\documentclass[aps,prd,amsfonts,amsmath,showpacs]{revtex4}
\usepackage[dvips]{graphicx}
\usepackage{bm}

\begin{document}

\title{Optimal Estimation of Non-Gaussianity}

\author{Daniel Babich}

\email{babich@physics.harvard.edu}

\affiliation{Deparment of Physics, Harvard University, Cambridge, MA 02138}

\affiliation{Harvard-Smithsonian Center for Astrophysics, 60 Garden Street, Cambridge, MA 02138}

\date{\today.}


\begin{abstract}  	
We systematically analyze the primordial non-Gaussianity estimator used by the
Wilkinson Microwave Anisotropy Probe (WMAP) science team with the basic ideas of 
estimation theory in order to see if the limited Cosmic Microwave Background (CMB) 
data is being optimally utilized. The WMAP estimator is based on the implicit 
assumption that the CMB bispectrum, the harmonic transform of the three-point 
correlation function, contains all of the primordial non-Gaussianity information 
in a CMB map. We first demonstrate that the Signal-to-Noise $(S/N)$ of an estimator 
based on CMB three-point correlation functions is significantly larger than the 
$S/N$ of any estimator based on higher-order correlation functions; justifying 
our choice to focus on the three-point correlation function.
We then conclude that the estimator based on the three-point correlation function,
which was used by WMAP, is optimal, meaning it saturates the Cramer-Rao Inequality 
when the underlying CMB map is nearly Gaussian. 
We quantify this restriction by demonstrating that the suboptimal character of 
our estimator is proportonal to the square of the fiducial non-Gaussianity, which is already 
constrained to be extremely small, so we can consider the WMAP estimator to be optimal 
in practice. Our conclusions do not depend on the form of the primordial 
bispectrum, only on the observationally established weak levels of primordial
non-Gaussianity.  
\end{abstract}

\maketitle

\section{Introduction}

The origin of the cosmological perturbations which led to the observed Cosmic Microwave Background (CMB) 
temperature anisotropies and large scale structure is one of the outstanding questions in cosmology. 
Unfortunately there are only a limited number of independent ways we can constrain the 
mechanism that produced these perturbations. Windows into the production mechanism of cosmological 
perturbations include a spectrum of primordial gravity waves, departures from a scale-invariant 
curvature power spectrum, isocurvature primordial perurbations and non-Gaussianity statistics of
the primordial curvature perturbations. While each of these features of the primordial 
perturbations will change the observed CMB anisotropies and provide insight into the production 
mechanism of the primordial perturbations, we will focus on the non-Gaussian characteristics of 
the CMB anisotropies in this paper. 

The standard theory of inflation robustly predicts that the primordial curvature perturbations, 
and therefore the resulting CMB anisotropies, should be nearly Gaussian \cite{Maldacena:2002vr}. 
By Gaussian we mean that all odd n-point correlation functions exactly vanish and all even n-point 
correlation functions can be completely expressed as the combination of two-point correlation functions. 
A non-Gaussian primordial curvature perturbation field would violate one of the above criteria. 
The degree to which any given Gaussianity criterion is violated can be characteristized by forming 
a natural ratio between the particular non-Gaussian correlation function and the appropriate combination 
of two-point corellation functions \cite{bernardeau}. Since we will ultimately perform calculations 
with a non-Gaussian Probability Distribution Function (PDF) (from which all n-point correlation 
functions can be calculated), below we will provide a natural quantity that describes the degree 
of non-Gaussianity in terms of quantites of the PDF. Then we can define the condition nearly Gaussian 
to mean that this natural quantity is much less than one. Even though the standard calculation within 
inflation predicts extremely small amounts of non-Gaussianity, there are several viable mechanisms 
that can produce substantial non-Gaussian primordial curvature perturbations (see \cite{Bartolo:2004if} 
and references there within). 
These processes not only predict different amplitudes of the total non-Gaussianity but also different 
functional forms \cite{Babich:2004gb}. Through linear gravitational and hydrodynamical evolution
these curvature perturbations will produce CMB anisotropies with statistical properties that
mirror the statistical properties of the primordial perturbations. Therefore it is possible
to learn about primordial non-Gaussianity by studying higher order n-point correlation functions
of the CMB anisotropies.

There are more potential sources of non-Gaussianity in the CMB than
just primordial non-Gaussianity. The non-linearities in the gravitational and hydrodynamical
equations of motion for the baryon-photon fluid prior to recombination can produce 
non-Gaussianity \cite{carroll,bartolo,Creminelli:2004pv}. Secondary anisotropies, such 
as, the thermal and kinetic Sunyaev-Zeldovich effects, gravitational lensing and the 
Ostriker-Vishniac effect can all produce non-Gaussianity in the 
CMB \cite{Goldberg:1999xm,Cooray:1999kg,Zaldarriaga:2000ud}. However we are
most interested in an observation of primordial non-Gaussianity in the CMB because this would
require new ideas for the production of the primordial curvature perturbations and may give a glimpse
into beyond the Standard Model physics. It has been demonstrated that the primordial non-Gaussian signal 
can be separated from non-Gaussian secondary anisotropies on scales relevant for WMAP and Planck 
\cite{Komatsu}.

As stated above we expect the non-Gaussianity of the CMB to be extremely small. Instead of trying to 
detect individual modes of a non-Gaussian correlation function \footnote{By individual mode, we mean
the correlation function evaluated for a particular configuration of Fourier wavevectors}, which would 
allow us to the examine the functional dependence of momentum wavevectors and better understand the 
mechanisms that produced the non-Gaussianity \cite{Babich:2004gb}, it is customary to parameterize 
the primordial non-Gaussianity with a model. Thus we will be able to combine many bispectra modes 
and therefore increase the statistical significance of our conclusions.
The standard model for primordial non-Gaussianity is the ``Local Model''; the primordial 
curvature perturbations in real-space are expressed as
\begin{equation}\label{local}
   \Psi({\bm x}) = \Phi({\bm x}) + f_{NL}[\Phi({\bm x})^2 - \langle \Phi({\bm x})^2\rangle],
\end{equation}
where $\Phi(\bm{x})$ is a Gaussian field with zero mean and covariance matrix $\mathcal{C}$. 
The ``Local Model,'' even though an idealization, has a strong physical motivation; during inflation
the non-linear couplings of general relativity will produce ``Local Model'' terms in the bispectrum, 
the harmonic transform of the connected three-point function \cite{Maldacena:2002vr}. Standard 
measures of the similarity of two bispectra imply that the ``Local Model''and the standard 
inflationary calculation are nearly identical \cite{Babich:2004gb}. 
Moreover, models where non-linearities develop outside the horizon, such as the curvaton 
model \cite{Lyth:2002my} and the inhomogeneous reheating models \cite{Dvali:2003em,Dvali:2003ar},
will produce bispectra identical to the ``Local Model.'' 
This model will allow us to place limits on the cumulative 
amplitude of all bispectrum modes and a method has been developed to translate these constraints 
into constraints on other models of primordial non-Gaussianity \cite{Babich:2004gb}. 

The characteristic amplitude of non-Gaussianity can be found by rescaling $\Psi$ to have unit 
variance. The coefficient of the rescaled quadratic term, $f_{NL} \langle \Phi({\bm x})^2\rangle^{1/2}$, 
is the natural measure of the amplitude of non-Gaussianity. Below we will find that the Signal-to-Noise
of estimators constructed out of non-Gaussian three-point correlation functions will always contain 
this factor (or related quantities for higher order non-Gaussian n-point correlation functions).
The best current data constrains 
$f_{NL} \langle \Phi({\bm x})^2\rangle^{1/2} \le 3.5 \times 10^{-3}$ ($95 \%$ C.L.) \cite{Komatsu:2003fd}.
The smallness of this expansion parameter not only allows us to truncate the expansion of the
definition of the non-Gaussian field, Eq. (\ref{local}), but it also implies that the characteristic 
amplitude of the bispectrum will be much larger than the characteristic amplitude of the trispectrum, 
the harmonic transform of the connected four-point function. 
By rescaling arguments analogous to above, the characteristic amplitude of the trispectrum, either due to 
an additional cubic term ($f_3 \Phi(\bm{x})^3$) or from the quadratic piece ($f_{NL} \Phi(\bm{x})^2 $), 
would be proportional to $(f^2_{NL} + f_3) \langle \Phi({\bm x})^2\rangle$. So 
unless $f_3$ is much larger than $f_{NL}$, the characteristic amplitude of the bispectrum will be 
much larger than the characteristic amplitude of the trispectrum. 

We are not simply interested in the characteristic amplitude of a higher-order correlation 
function, but in the cummulative Signal-to-Noise $(S/N)$ of an estimator based on that 
higher-order correlation function.
The number of relevant graphs grows quickly as the order of the correlation function increases, for 
example the number of trispectrum quadrilaterals is much greater than the number of bispectrum triangles.
As demonstrated in Appendix A, if the primordial power spectrum of the Gaussian field is scale-invariant, 
then the $(S/N)^2$ of a non-Gaussianity estimator based on any non-Gaussian correlation function, 
regardless of order, will simply scale with the number of observed pixels ($N_{pix}$), not the number of graphs. 
Ignoring combinatorial factors we find, for an estimator constructed out of the $n^{th}$-point correlation 
function, that 
\begin{equation}
(S/N)_n \sim (f_{n-1} + f^2_{n-2}+\cdots + f^{n-2}_2) \langle \Phi({\bm x})^2\rangle^{(n-2)/2} \sqrt{N_{pix}}. 
\end{equation}
Since $\langle \Phi({\bm x})^2\rangle^{1/2} \sim 10^{-5}$ the estimator based on the three-point correlation 
function ($f_{NL} \equiv f_2$) will dominate over all higher-order estimators. 
This implies that the three-point correlation function is the most effective 
means to constrain $f_{NL}$ and that it will be significantly easier to constrain $f_{NL}$ than
any higher-order $f_n$. Thus we should begin to explore primordial non-Gaussianity by constraining 
the amplitude of the bispectrum.

These conclusions depend upon the naturalness argument that $f_{NL} \sim f_{3} \sim f_{n}$, which 
implies the $S/N$ of the various estimators will be supressed by increasing factors of 
$\langle \Phi({\bm x})^2\rangle^{1/2}$, which we believe to be approximately $10^{-5}$. 
There are several reasons to be cautious.
First of all, there might be some reason due to symmetry that $f_{NL} = 0$, then the primordial
bispectrum will vanish. 
Also in the inhomogeneous reheating models, inefficiency in the production of the gravitational
potential could equally supresses all terms in expansion of the non-Gaussian field; therefore
Eq. (\ref{local}) would become
\begin{equation}
	\Psi({\bm x}) = A[\Phi({\bm x}) + f_{NL}[\Phi({\bm x})^2 - \langle \Phi({\bm x})^2\rangle]],   
\end{equation} 
where $A < 1$ \cite{Matias}. This mutual suppression would increase the inferred value of 
 $\langle \Phi({\bm x})^2\rangle^{1/2}$ and cause the suppression of the $S/N$ of higher-order 
estimators to be less drastic. However within the standard slow-role single field inflationary 
scenario \cite{Maldacena:2002vr} our conclusions are quite robust.

The standard inflationary scenario for the generation of primordial curvature perturbations predicts that
these perturbations were quantum mechanically produced. While it is impossible for us to predict
the primordial curvature fluctuation at any point in the universe, it is possible to predict the 
statistical properties of these perturbations. With these statistical properties we can create a 
PDF from which we can generate random realizations of the 
non-Gaussian curvature perturbations. Since the CMB anisotropies were generated through the 
linear gravitational and hydrodynamical evolution of these primordial curvature perturbations,
the statistical properties of the CMB anisotropies will mirror the statistical properties of
the primordial curvature perturbations.
Therefore we can view the data from a given CMB experiment as random samples drawn from the appropriate
CMB PDF derived from the Local Model PDF.
 
We have an explicit expression for the primordial curvature perturbations, Eq. (\ref{local}), 
so we can write down the exact PDF for this random field 
\begin{equation}\label{curv_dist}
   P(\Psi | f_{NL}, \mathcal{C}) = \int d^N \Phi \delta^{(N)} 
       (\Psi({\bm x}) - \Phi({\bm x})-f_{NL}[\Phi({\bm x})^2 - \langle \Phi({\bm x})^2\rangle]) 
      P(\Phi | \mathcal{C}),
\end{equation}
and we also include the assumption that $\Phi$ has a Gaussian distribution
\begin{equation}
   P(\Phi| \mathcal{C}) = \frac{e^{- \Phi^T \mathcal{C}^{-1} \Phi/2}}{\sqrt{(2\pi)^N\det{\mathcal{C}}}}. 
\end{equation} 
Here we have adopted standard notation for functionals and compactly written the quadratic form of 
functionals in vector notation. Depending on the particular situation, the quantity 
$\Phi^T \mathcal{C}^{-1} \Phi$ could either be a sum over a discrete set of eigenfunctions 
(for example, observations of CMB anisotropies $\Phi^T \mathcal{C}^{-1} \Phi = \sum_{lm} a_{lm}^* a_{lm}/C_l$ )
or an integral over a continuous set of eigenfunctions (for example, observations of primordial curvature 
perturbations $\Phi^T \mathcal{C}^{-1} \Phi = \int d^3\bm{k} \Phi^*(\bm{k}) \Phi(\bm{k})/P(k)$).
Assuming that the Local Model is correct and the PDF in Eq. (\ref{curv_dist}) properly represents nature, 
we can ask how well the data sample will allow us to constrain the underlying parameters of the PDF 
($f_{NL}, \mathcal{C}$). There are standard tools from the field of statistical estimation theory 
that will help us rate estimation procedures and determine the smallest possible error bars we 
can place on a parameter with a given set of data \cite{Kay,Zacks,Poor,Kullback,Kendall}. 

The purpose of the paper is to determine what are the smallest error bar possible to place on $f_{NL}$
with a given data set and which estimation procedure will allow us to place these constraints. The 
entire set of measured n-point correlation functions contains all observable information on the underlying 
PDF contained in the given data set. It may be possible to find a simple estimator which retains all of
the available information without undergoing the time consuming process of measuring all n-point correlation
functions. Currently researchers are using several different techniques to estimate the non-Gaussianity 
of a CMB map. An estimator based on the CMB three-point correlation function has been used 
to constrain the non-Gaussianity ($f_{NL}$) of the results from the Wilkinson Microwave Anisotropy Probe (WMAP) 
\cite{Komatsu:2003fd, Komatsu:2003iq}, Very Small Array (VSA) \cite{Smith:2004jd} and 
MAXIMA \cite{Santos:2002df} CMB experiments. In addition, Minkowski Functionals \cite{Eriksen:2004df} 
and the correlations of Fourier phases (see \cite{Chiang:2004yg,Chiang:2002vm} and references within) 
have been used to characterize non-Gaussianity. We will demonstrate, that for weak 
levels of non-Gaussianity, the estimator based on the three-point function contains all possible information
on $f_{NL}$. In particular we will show that the exact estimator used by WMAP is an optimal
estimator. Therefore the procedure adopted by WMAP will, in principle, provide the best possible error
bars on $f_{NL}$. By using this estimator none of the information potentially contained in the higher
order n-point functions is lost. This conclusion is equivalent to the statement that an estimator
based on the two-point correlation function contains all possible information on the power spectrum, even
though higher order even n-point correlations functions also contain information on the power spectrum. 
Finally we will argue that our conclusion does not depend on any characteristic of the Local Model, 
just on the observationally established weak levels of non-Gaussianity, so the amplitude of any primordial 
bispectrum can be optimally constrained by estimators based on the CMB three-point correlation function.

The calculations in this paper assume that the CMB map is free of both instrument noise and Galactic foreground
contamination. It is straightforward to include the effects of noise in our estimator for $f_{NL}$. However
the statistical estimation of individual bispectrum modes is very complicated for maps with inhomogeneous
noise or a Galactic sky cut. The simple calculation of individual bispectrum modes by combining the 
appropriate coefficient of the spherical harmonic decomposition of the observed CMB map will result in
sub-optimal estimates for the bispectrum modes and therefore sub-optimal estimates for $f_{NL}$. An
optimal estimator for individual bispectrum modes, including the effects of inhomogeneous instrument noise,
has been developed \cite{Heavens:1998jb}. In this paper we treat the part of the problem which involves
the estimation of $f_{NL}$ from measured bispectrum modes.

In Section II we introduce the notion of an optimal estimator and the Cramer-Rao Inequality 
and show how they are related to the Maximum Likelihood Estimator. In Section III we analyze a Poisson 
random field with the appropriate Local Model non-Gaussian PDF as a simple example of the essential ideas. 
In Section IV we extend our analysis to the scale-invariant distribution predicted by inflation and 
generalize our results to arbitrary primordial bispectrum in Section V. In Section VI we summarize our results.

\section{Estimation Theory}

Estimation theory is the branch of statistics developed in order to analyze the procedures used
to constrain the underlying continuous parameters of a PDF. Assuming that a given data set is drawn
from a known PDF with unknown fixed parameters it allows us to determine the minimum error bars
attainable on those parameters using the data. In the following subsections we develop a 
simple sufficient and necessary condition for a PDF to admit an estimator that saturates the 
famous Cramer-Rao Inequality. 

Then for completeness we will discuss the relationship between 
these concepts and the popular Maximum Likelihood Estimator (MLE). We will only outline necessary
concepts, the interested reader should consult \cite{Kay,Zacks,Poor,Kullback,Kendall} for more
details.

\subsection{Cramer-Rao Inequality}

In this subsection we discuss the Cramer-Rao Inequality, which determines the lower bound
for error bars that we can place on a parameter with a given data sample. These minimum 
errors bars derived from the Cramer-Rao Inequality are valid only for the ``frequentist,''
and not the Bayesian, understanding of statistical estimation (see 
\cite{Kay,Zacks,Poor,Kullback,Kendall} for a discussion of these differing understandings 
of probability and estimation). An estimator that 
saturates this inequality is dubbed optimal since it weights the data in the best possible 
manner. Unfortunately for a general PDF no optimal estimator will exist.
In order to develop some intuition regarding the information content of a data sample 
we will derive the Cramer-Rao Inequality for a scalar parameter. The generalization
to multiple parameters is conceptly straightforward, but requires linear algebra
which obscures some of the essential insights.

We let $\Lambda(\bm{x})$ be an unbiased estimator of $\lambda$, the parameter we
are trying to estimate from our data sample. Assuming the regularity condition
\begin{equation}\label{reg_cond}
   \int \frac{\partial \ln{p(\bm{x}|\lambda)}}{\partial \lambda} p(\bm{x}|\lambda) d\bm{x} = 0,
\end{equation}
which holds when we can interchange the order of integration and differentiation, we can 
use this property to derive the following identity
\begin{equation}\label{unbiased}
   \int (\Lambda(\bm{x}) - \lambda) \frac{\partial \ln{p(\bm{x}|\lambda)}}{\partial \lambda} p(\bm{x}|\lambda) d\bm{x} = 1.
\end{equation}
Using the Schwarz Inequality, $(\bm{A}\cdot\bm{B})^2 \le (\bm{A}\cdot\bm{A})(\bm{B}\cdot\bm{B})$ in the notation 
of linear algebra, we find
\begin{equation}\label{Schwarz}
   [\int (\Lambda(\bm{x}) - \lambda) \frac{\partial \ln{p(\bm{x}|\lambda)}}{\partial \lambda} p(\bm{x}|\lambda) d\bm{x}]^2 \le
	[\int (\Lambda(\bm{x}) - \lambda)^2 p(\bm{x}|\lambda) d\bm{x}] 
        [\int (\frac{\partial \ln{p(\bm{x}|\lambda)}}{\partial \lambda})^2 p(\bm{x}|\lambda) d\bm{x}].
\end{equation}
Now defining the Fisher Information as
\begin{equation}\label{def_fisher}
   F(\lambda) = \int (\frac{\partial \ln{p(\bm{x}|\lambda)}}{\partial \lambda})^2 p(\bm{x}|\lambda) d\bm{x},
\end{equation}  
and using Eq. (\ref{unbiased}) we obtain the Cramer-Rao Inequality
\begin{equation}\label{CR_Ineq}
   Var(\Lambda(\bm{x})) \equiv [\int (\Lambda(\bm{x}) - \lambda)^2 p(\bm{x}|\lambda) d\bm{x}] \ge \frac{1}{F(\lambda)}.
\end{equation}
The Cramer-Rao Inequality states that no estimator of $\lambda$ can produce error bars smaller than 
$1/F(\lambda)$.

More importantly we can identify a necessary and sufficient condition for the variance of $\Lambda(\bm{x})$ 
to saturate the Cramer-Rao Inequality. It is clear that the Schwarz Inequality in Eq. (\ref{Schwarz}) will 
be saturated if and only if
\begin{equation}\label{nec_suf}
   \frac{\partial \ln{p(\bm{x}|\lambda)}}{\partial \lambda} = F(\lambda) (\Lambda(\bm{x}) - \lambda),
\end{equation}
thus the PDF must be able to be written in this form if it will admit an optimal estimator. 
In what follows, we will analyze the Local Model PDF to determine if it can be expressed in this form.

There is a multiparameter generalization of the Fisher Information defined in Eq. (\ref{def_fisher}),
\begin{equation}
  \mathcal{F}_{ij} = \int \frac{\partial \ln{p(\bm{x}|\lambda)}}{\partial \lambda_i} 
      \frac{\partial \ln{p(\bm{x}|\lambda)}}{\partial \lambda_j}  p(\bm{x}|\lambda) d\bm{x}.
\end{equation}
If we define $\mathcal{C}_{ij}$ as the covariance matrix of the set of parameters, the Cramer-Rao Inequality
becomes the statement that $\mathcal{C}-\mathcal{F}$ is a positive semidefinite matrix. This implies
the more familiar statement
\begin{equation}\label{multi_fisher}
  Var(\lambda_i) \equiv \mathcal{C}_{ii} \ge \mathcal{F}^{-1}_{ii}.
\end{equation}
If all off diagonal terms in $\mathcal{C}_{ij}$ vanish then we can independently estimate all $\lambda_i$
and the multiparmeter Cramer-Rao Inequality reduces the single parameter case.

\subsection{Relationship with the Maximum Likelihood Estimator}
Previous work on the analysis of data focused on the Maximum Likelihood Estimator (MLE). 
Here we will discuss the relationship between the more familiar MLE and the optimal estimator 
that we introduced above. 
The basic principle of the MLE is to consider the observed data as fixed and consequentially 
choose $\lambda$ in order to maximize the probability of observing the fixed data. When we 
take this approach the PDF will be refered to as the likelihood function. We must choose 
$\lambda$ in order to maximize the likelihood function, so equivalently we demand
\begin{equation}\label{mle_def}
   \frac{\partial \ln{p(\bm{x}|\lambda)}}{\partial \lambda}|_{\lambda=\lambda_{ML}} = 0, 
\end{equation}
then $\lambda_{ML}$ is the estimated value of $\lambda$.
Notice that this equality is true for all data realizations and not just for ensemble
averages. In general Eq. (\ref{mle_def}) will be a complicated non-linear equation, but
there are standard techniques to solve such equations. The most popular is the 
Newton-Raphson method which is widely used in cosmology for likelihood estimation
of the CMB power spectrum from observed CMB maps \cite{Bond:1998zw,Oh:1998sr}.

This approach is widely used in practice because it is always possible to implement, 
whereas the approach described above often does not yield an estimator.  
In addition, the MLE is asymptotically optimal and unbiased, meaning that as the amount 
of data increases the MLE approaches the correct answer with error bars equal to those 
predicted by the Cramer-Rao Inequality. Whenever an optimal estimator exists, the 
optimal estimator is also the MLE. This is clear from the necessary and sufficient 
condition for the existence of an optimal estimator, Eq. (\ref{nec_suf}), which 
automatically satisfies the definition of the MLE, Eq. (\ref{mle_def}). However the 
converse does not hold and in general, for finite amounts of data, the MLE is not optimal. 

\section{Poisson Distributions}

In order to gain intuition we will first analyze a Poisson random field with the 
appropriate non-Gaussian PDF. This implies that each point in space is uncorrelated 
with one another and will be independently sampled from the non-Gaussian Local Model PDF. 
Since we measure N independent and identically distributed (IID) random variables we 
can simply scale the single pixel Fisher Matrix by N. With these assumptions, the PDF 
for the single pixel primordial curvature perturbation, Eq. (\ref{curv_dist}), becomes
\begin{equation}\label{PDF_def}
   P(\Psi | f_{NL}, \mu) = \int d \Phi \delta(\Psi-\Phi-f_{NL}[\Phi^2 - \mu^2] )
	\frac{e^{-\Phi^2/2\mu^2}}{\sqrt{2\pi \mu^2}}, 
\end{equation} 
where $\mu^2 = \langle \Phi^2 \rangle$.
Given an observed $\Psi$ there are two possible values of the underlying $\Phi$
\begin{equation}
\Phi_{\pm} = \frac{1}{2 f_{NL}} [\pm \sqrt{1+4f_{NL} (\Psi + f_{NL} \mu^2)} -1].
\end{equation}
Integrating over the delta-function the PDF becomes
\begin{equation}
 P(\Psi | f_{NL}, \mu) = \frac{1}{\sqrt{2\pi \mu^2}} [ \frac{e^{-\Phi_+^2 / 2\mu^2}}
  {1+2f_{NL}\Phi_+} +  \frac{e^{-\Phi_-^2 / 2\mu^2}}{1+2f_{NL}\Phi_-} ].
\end{equation}

In the weakly non-Gaussian limit, $f_{NL}\mu \ll 1$, the contribution from $\Phi_-$ is exponentially suppressed so 
we will ignore it in what follows; we can then write
\begin{equation}\label{log_PDF}
\log P(\Psi | f_{NL}, \mu) = - \frac{\Phi_+^2}{2 \mu^2} 
- \ln (1+2f_{NL}\Phi_+ ) - \ln \mu,
\end{equation}
which can be expanded in a power series
\begin{equation}\label{poisson_ps}
\log P(\Psi | f_{NL}, \mu) = - \frac{\Psi^2 + 2 \mu^2 \ln \mu^2}{2 \mu^2} + \frac{f_{NL}}{\mu^2}(\Psi^3- 3 \mu^2 \Psi) 
- \frac{f^2_{NL}}{2\mu^2}(5 \Psi^4 + 5 \mu^4 - 14\mu^2\Psi^2) + \mathcal{O}(f_{NL}^3).
\end{equation}
If we rewrite $\log P$ such that the power series is expressed in terms
of $\Psi/\mu$, which is a random variable with unit variance, then we discover that we are 
actually expanding in the quantity $f_{NL} \mu$. Doing so we can write Eq. (\ref{poisson_ps})
as
\begin{equation}
  \log P(\Psi | f_{NL}, \mu) = -I_0(\Psi/\mu) + f_{NL}\mu I_1(\Psi/\mu) - \frac{1}{2} (f_{NL}\mu)^2 I_2(\Psi/\mu)
  + \mathcal{O}(f^3_{NL}\mu^3),
\end{equation}
where $I_0, I_1$ and $I_2$ are defined with respect to Eq. (\ref{poisson_ps}).
The expectation value of the first-order piece is 
\begin{equation}
  \langle I_1 \rangle = 6 f_{NL} \mu + \mathcal{O}(f^3_{NL} \mu^3),
\end{equation}
and the second-order piece is
\begin{equation}
  \langle I_2 \rangle = 6 + 272 f^2_{NL} \mu^2 + \mathcal{O}(f^4_{NL} \mu^4).
\end{equation}
Thus it is necessary to keep the second order term, $I_2$, in the expansion of $\log P$. However we can 
ignore the non-Gaussian piece of $\langle I_2 \rangle$, which is a factor of $f^2_{NL} \mu^2$ smaller 
than the Gaussian piece of $\langle I_2 \rangle$. Our conclusions will depend on this approximation, 
which is equivalent to the weakly non-Gaussian approximation typically made in the literature.

If we view the PDF as a likelihood function, as described above, only $I_1$ contains 
information on $f_{NL}$ to $\mathcal{O}(f^4_{NL} \mu^4)$. Since we only need to
calculate $I_1$ from the observed data to specify the likelihood we can regard this
quantity as an estimator for $f_{NL}$; we will address the need to specify $I_2$ below.
Now we will analyze the statistical properties of this estimator.

After choosing the normalization in order to unbias the estimator we find
\begin{equation}\label{est_f}
  \hat{f}_{NL} = \frac{1}{6\mu^4}(\Psi^3-3\mu^2\Psi).
\end{equation}
Once we include all N independent observations, this estimator becomes
\begin{equation}\label{MVUE}
  \hat{f}_{NL} = \frac{1}{6\mu^4N}\sum_i^N(\Psi_i^3-3\mu^2\Psi_i),
\end{equation}
which has the variance
\begin{equation}\label{est_var}
  Var(\hat{f}_{NL}) = \frac{1}{6\mu^2N} + \frac{22 f^2_{NL}}{N} + \mathcal{O}(f^4_{NL}\mu^4).
\end{equation}

The PDF defined in Eq. (\ref{PDF_def}) satisfies the regularity condition, Eq. (\ref{reg_cond}), 
for both $f_{NL}$ and $\mu$ so we can use the Cramer-Rao Inequality to give the lower bound on 
the error bars of these parameters. This lower bound will be different than the variance of 
$\hat{f}_{NL}$ if the PDF does not satisfy the appropriate necessary and sufficient condition for 
a PDF to admit an optimal estimator, Eq. (\ref{nec_suf}). Simple inspection of the single 
pixel PDF, Eq. (\ref{PDF_def})
\begin{equation}\label{poisson_lgpdf}
  \frac{\partial \ln{p(\Psi|f_{NL})}}{\partial f_{NL}} = 6\mu^2[\hat{f}_{NL}(\Psi)
   - \frac{f_{NL}}{6} I_2(\Psi/\mu)] + \mathcal{O}(f^2_{NL}\mu^2),
\end{equation}
shows that the necessary and sufficient condition of Eq. (\ref{nec_suf}) is strictly met only 
if $f_{NL} = 0$. Thus our estimator $\hat{f}_{NL}(\Psi)$ is only optimal for setting 
non-Gaussian limits on Gaussian maps.

The PDF does not satisfy the appropriate conditions for the existance of an optimal estimator because
the function $I_2(\Psi/\mu)$ multiplies $f_{NL}$. When we evaluate $\langle I_2 \rangle$, we 
find that there is a leading order Gaussian piece and a non-Gaussian piece, suppressed by a factor 
of $f^2_{NL} \mu^2$. If we replace $I_2$ with its expectation value, ignoring the non-Gaussian piece, 
we find
\begin{equation}
  \frac{\partial \ln{p(\Psi|f_{NL})}}{\partial f_{NL}} = 6\mu^2[\hat{f}_{NL}(\Psi) - f_{NL}],
\end{equation}
which is exactly the condition for an optimal estimator to exist. While our original conclusion, 
that $\Psi^3 -3 \mu^2 \Psi$ is only optimal for underlying Gaussian distributions, is still true; 
we can clearly see that within the weakly non-Gaussian limit our estimator is optimal. If the 
estimator was optimal its variance would exactly equal the bound derived from the Fisher Matrix. 
For the non-Gaussian underlying distribution (the inverse of) the Fisher Matrix, Eq. 
(\ref{poisson_lgpdf}), and the estimator variance, Eq. (\ref{est_var}), differ by terms proportional 
to $f^2_{NL} \mu^2$. This is precisely the type
of term that can be ignored within the weakly non-Gaussian approximation. Therefore assuming this
approximation is valid, the estimator $\Psi^3 -3 \mu^2 \Psi$ is optimal even for underlying 
non-Gaussian distributions.

This is not coincidence, but a direct consequence of the regularity condition, Eq. (\ref{reg_cond}), 
and the weakly non-Gaussian approximation. Since the regularity condition
must hold for all values of $f_{NL}$, we can require it to hold term-by-term in our expansion
in $f_{NL}$. The first-order terms in the regularity condition are 
$\langle I_1 \rangle_{NG} - f_{NL} \langle I_2 \rangle_{G}$, where G and NG denote the Gaussian 
and non-Gaussian expectation values, respectively. 
Since we know the first-order terms must vanish, we can infer $\langle I_1 \rangle_{NG} 
= f_{NL} \langle I_2 \rangle_{G}$. Thus if we can replace $I_2$ with its Gaussian expectation
value then $\hat{f}_{NL} = I_1/\langle I_2 \rangle_{G}$ will be an unbiased optimal estimator 
with variance $1/\langle I_2 \rangle_{G}$.

Our argument depends on being able to replace $I_2$ with its expectation value, clearly when the 
sample size grows the error introduced by making this assumption vanishes. Now we will calculate 
conditions on the sample size for this property to hold. The expectation value will have a variance 
which decreases inversely with the sample size, $N$,
\begin{equation}
  Var(\frac{I_2(\Psi/\mu)}{6}) = \frac{287}{9N},
\end{equation}
thus we need $N \gg 287/9$ for our estimator to be optimal. This number is simply due to combinatorial
factors and therefore we do not expect it to drastically change when we consider the full problem
with radiative transfer and a scale-invariant distribution. Moreover WMAP observes approximately
$10^6$ pixels, so we expect this approximation to be quite good.

To check our conclusions, we should still calculate the Fisher Matrix elements and compare them 
with the variance of $\hat{f}_{NL}$. After scaling our results, since we have N IID samples, we 
find to $\mathcal{O}(f^3_{NL} \mu^3)$
\begin{equation}\label{fisher_matrix}
F=N \left(\begin{array}{cc} 6\mu^2 + 368 f^2_{NL} \mu^4 & -8f_{NL}\mu \\
			      -8f_{NL}\mu & 2/\mu^2 + 20f^2_{NL}  \\
\end{array} \right), 
\end{equation}
where $\bm{\lambda} = (f_{NL}, \mu)$.
The errors on the parameters are given by the inverse of the Fisher Matrix 
\begin{equation}
F^{-1}= \frac{1}{N} \left(\begin{array}{cc} 
					     1/ 6\mu^2 - 28f^2_{NL}/3 & 2f_{NL} \mu/3 \\
				             2f_{NL} \mu/3 & \mu^2/2 - 7f^2_{NL}\mu^4/3 \\ 
\end{array} \right), 
\end{equation}
thus the Cramer-Rao bound on $f_{NL}$ is smaller than the variance of $\hat{f}_{NL}$ except when our
underlying map is Gaussian. 

The fractional discrepancy caused by underlying non-Gaussianity is
\begin{equation}
  \frac{Var(\hat{f}_{NL})-(F_{ff})^{-1}}{Var(\hat{f}_{NL})} = 170 f^2_{NL} \mu^2 \le 2.3 \times 10^{-3}
\end{equation}
assuming, $f_{NL}\mu \le 3.5 \times 10^{-3}$, which is the best current constraint. Clearly we 
are justified in ignoring the $\mathcal{O}(f^2_{NL}\mu^2)$ corrections and in considering the 
three-point function to be optimal even when the underlying non-Gaussianity is non-zero.

We can extend this discussion by removing the assumption that $\mu$ is known {\it a priori}, 
then we must simultaneously estimate $f_{NL}$ and $\mu$. Dropping this assumption we will 
search for an estimator of $f_{NL}$ that jointly estimates $\mu$. Actually 
it is most natural to estimate $\lambda = f_{NL} \mu^4$, using this variable we find that the 
coupled unbiased estimator is
\begin{equation}
   \hat{\lambda} = \frac{1}{6} \frac{N}{N-3} (\frac{1}{N}\sum_i \Psi^3_i - \frac{3}{N^2}\sum_{ij} \Psi^2_i \Psi_j),
\end{equation}
with variance
\begin{equation}
  Var(\hat{\lambda}) = \frac{\mu^6}{6 N}\frac{N^2 - 3N + 12}{(N-3)^2}.
\end{equation}
At $N = 100$, the variance of $\hat{\lambda}$ is increased by just $3 \%$ with respect to $\hat{f}_{NL}$. 
In general the simultaneous estimation of parameters that are correlated will increase the
estimator variance. However the variance will asymptotically approach the single parameter estimator
variance as the size of the data sample grows. 

\section{Scale-Invariant Local Model non-Gaussian Distributions}

The purpose of this section is to determine whether an optimal estimator for $f_{NL}$ exists when 
we change the underlying distribution from Poisson to scale-invariant. 
Observations of the CMB and large scale structure imply that the primordial spectrum of curvature
perturbations is scale-invariant; therefore, we must adapt the discussion of the previous
section. The Poisson distribution covariance matrix is diagonal in all bases; however, the real-space
basis is convenient because the ``Local Model'' definition of the non-Gaussian field does not 
mix different modes in this basis. Therefore in the real-space basis we were able to analyze a 
single pixel and appropriately scale the final results.
For a scale-invariant distribution the covariance matrix is no longer diagonal in a real-space basis, 
but only in a Fourier basis.
Moreover we observe the primordial curvature perturbations projected on the sky as CMB anisotropies 
after hydrodynamical and gravitational evolution. The covariance matrix of the CMB anisotropies
is diagonal in a spherical harmonic basis. 

Taking these features into account we will determine if the bispectrum, the three-point correlation 
function in the spherical harmonic basis, contains all of the non-Gaussian information of a CMB map. 
This problem will be divided into two parts: 
(1) We will analyze a 2-D scale-invariant distribution without radiative transfer in a spherical 
harmonic basis;
(2) We will include the effects of radiative transfer and show that the WMAP non-Gaussianity estimator 
is an optimal estimator for $f_{NL}$.

\subsection{Sachs-Wolfe Effect}

At first we will ignore the effects of radiative transfer and consider scale-invariant primordial curvature
perturbations projected onto the sky. This essentially occurs on large angular scales
where the Sachs-Wolfe effect directly maps the curvature perturbations onto temperature anisotropies. 
However, the main purpose of this subsection is simply to show the changes in the PDF as we switch from an 
underlying Poisson distribution to a scale-invariant one.

The two-point correlation function of the projected Gaussian curvature field is
\begin{equation}
   \mathcal{C} \equiv \langle \Phi^{*}_{lm} \Phi_{l' m'} \rangle = \delta_{l,l'}\delta_{m,m'}
   \frac{2}{\pi} \int k^2 dk P(k) \frac{j^2_l(k\tau_D)}{9},
\end{equation}
where $\tau_D$ is the distance to the surface of last scattering. For a scale-invariant power 
spectrum, $P(k) = A/k^3$, can be exactly evaluated as
\begin{equation}
   \langle \Phi^{*}_{lm} \Phi_{l' m'} \rangle \equiv \delta_{l,l'}\delta_{m,m'} D_l 
   = \delta_{l,l'}\delta_{m,m'} \frac{A}{9 \pi}\frac{1}{l(l+1)}.
\end{equation}

Now defining the Gaunt Integral as
\begin{eqnarray}\label{Gaunt}
   \mathcal{G}^{l_1 l_2 l_3}_{m_1 m_2 m_3} &=& 
      \int d^2\hat{n} Y_{l_1 m_1}(\hat{n}) Y_{l_2 m_2}(\hat{n}) Y_{l_3 m_3}(\hat{n}) \\
  &=& \sqrt{\frac{(2l_1+1)(2l_2+1)(2l_3+1)}{4\pi}} 
   \left(\begin{array}{ccc} l_1 & l_2 & l_3 \\ 0 & 0 & 0 \end{array}\right)
   \left(\begin{array}{ccc} l_1 & l_2 & l_3 \\ m_1 & m_2 & m_3 \end{array}\right),
\end{eqnarray}
and the average of the quadratic curvature fluctuation as
\begin{equation}
   \langle \Phi^2(\bm{x}) \rangle = \mu^2 = \int \frac{d^3\bm{k}}{(2\pi)^3} P(k),
\end{equation}
we can follow the above steps to expand the non-Gaussian PDF in powers of $f_{NL} \mu$. 
However, we must first calculate the Jacobian transformation of the delta-function
\begin{eqnarray}
   J_{lm,ab} & \equiv & \frac{\partial g_{lm}}{\partial \Phi_{ab}}
	= - \delta_{l,a} \delta_{m,b} - f_{NL} (-1)^{m} \sum_{l_1,l_2,m_1,m_2} 
       \mathcal{G}^{l l_1 l_2}_{m m_1 m_2}\frac{\partial}{\partial \Phi_{a,b}} 
       \Phi_{l_1 m_1} \Phi_{l_2 m_2}, \\
        & = & - \delta_{l,a} \delta_{m,b} - 2f_{NL} (-1)^{m} \sum_{l_1,m_1} 
	\mathcal{G}^{l l_1 a}_{-m m_1 b} \Psi_{l_1 m_1} + \mathcal{O}(f^2_{NL}\mu^2),
\end{eqnarray}
where $g_{l m}$ is the argument of the delta-function, used to define the Local Model PDF
in Eq. (\ref{curv_dist}), expressed in a spherical harmonic basis.

For the Local Model non-Gaussian PDF we find
\begin{equation}\label{logPDF_2d}
   \log{P(\Psi|f_{NL},\mathcal{C})} = -\frac{\Psi^T \mathcal{C}^{-1} \Psi}{2} 
     + f_{NL} \Psi^T \mathcal{C}^{-1}(\Psi^2-\mu^2) - \log{\det{\frac{\partial g(\Phi)}{\partial \Phi}}} 
     + \mathcal{O}(f^2_{NL}\mu^2),
\end{equation}
which expressed in a spherical harmonic basis is
\begin{eqnarray}
  \log{P(\Psi|f_{NL},\mathcal{C})} &=& -\frac{1}{2} \sum_{l m} \frac{\Psi^{*}_{l m} \Psi_{l m}}{D_l} + 
  f_{NL}\sum_{(l,m)} \mathcal{G}^{l_1 l_2 l_3}_{m_1 m_2 m_3} \frac{\Psi_{l_1 m_1}}{D_{l_1}}\Psi_{l_2 m_2}\Psi_{l_3 m_3}
  - f_{NL}\frac{\mu^2\Psi_{0 0}}{D_0} \nonumber \\
 &-& \log{\det{\frac{\partial g(\Phi)}{\partial \Phi}}} + \mathcal{O}(f^2_{NL}\mu^2).
\end{eqnarray}
The notation $(l,m)$ is meant to imply that the sum is over all three $l_i$ and $m_i$.
It is clear that rotational invariance, enforced through the Gaunt Integral selection rules, 
forces the terms linear in $\Psi$ to only contain the monopole term, $\Psi_{0 0}$. 
	
We can simplify matters by examining the derivative of the expression in Eq. (\ref{logPDF_2d}); 
using the standard matrix identity, Log~Det~A = Tr~Log~A, to rewrite the term that comes from the 
Jacobian transformation of the delta-function, we find
\begin{equation}
   \frac{\partial \log{P(\Psi|f_{NL},\mathcal{C})}}{\partial f_{NL}} = \Psi^T \mathcal{C}^{-1}(\Psi^2-\mu^2) 
   - Tr[J^{-1}\frac{\partial J}{\partial f_{NL}}] + \mathcal{O}(f_{NL}\mu),    
\end{equation}
which expressed in the spherical harmonic basis is
\begin{equation}\label{SI_suff}
   \frac{\partial \log{P(\Psi|f_{NL},\mathcal{C})}}{\partial f_{NL}} 
    = \sum_{(l,m)} \mathcal{G}^{l_1 l_2 l_3}_{m_1 m_2 m_3} \frac{\Psi_{l_1 m_1}}{D_{l_1}} 
   \Psi_{l_2 m_2} \Psi_{l_3 m_3} - \frac{\mu^2 \Psi_{0 0}}{D_0} 
   - 2 \Psi_{0 0} + \mathcal{O}(f_{NL}\mu).
\end{equation}
Using our intuition from the Poisson case we can identify
\begin{equation}
f_{NL}(\Psi) = \sum_{(l,m)} \mathcal{G}^{l_1 l_2 l_3}_{m_1 m_2 m_3} \frac{\Psi_{l_1 m_1}}{D_{l_1}} \Psi_{l_2 m_2} 
  \Psi_{l_3 m_3} - \frac{\Psi_{0 0}}{D_0}(2\mu^2+D_0),
\end{equation}
as the scale-invariant generalization of the estimator for $f_{NL}$. 
There are notable differences between the scale-invariant case and the Poisson case. 
Fundamentally, these differences are a result of expressing the PDF in a spherical 
harmonic basis instead of a real-space basis. The assumption of scale-invariance only 
affects the form of $D_l$. 

Rotational invariance forces the linear terms to be proportional to the monopole anisotropy modes.
Recall that the CMB monopole anisotropy is unmeasurable in principle and the primordial 
dipole anisotropy is dominated by the Doppler effect due to the local kinematic motion of 
the galaxy, so these modes are typically removed from the data. Labelling the original data
as $\Psi^0$ and the new data with the monopole and dipole modes eliminated as $\Psi^1$, the 
projection operator will change the PDF of the original data as 
\begin{equation}
  P(\Psi^1) = \int d^N\Psi^0 \delta^{(N-4)}(\Psi^1 - \bm{M} \Psi^0) P(\Psi^0),
\end{equation}
where $\bm{M}$ is the relevant projection matrix. For all non-monopole and dipole modes the
effect of the functional integration will simply be to replace $\Psi^0$ with $\Psi^1$. Since
the projection matrix eliminates the monopole and dipole modes in the argument of the 
delta-function, the monopole and dipole terms are simply integrated out. This functional 
integration simply eliminates the term in $P(\Psi)$ which is linear in $\Psi_{0 0}$. 

Moreover this functional integration also removes all terms from the cubic piece that contain 
any $\Psi_{0 0}$ or $\Psi_{1 m}$ modes. It is clear that any term linear or cubic in
a monopole or dipole mode will vanish. However it is possible that terms such as
$\Psi_{0 0} \Psi_{0 0} \Psi_{l m}$ or $\Psi_{1 1} \Psi_{1 -1} \Psi_{l m}$, which survive the
functional integration, might introduce a new linear term containing higher modes. 
Fortunately the symmetry properties of the Gaunt Integral eliminate these terms since
$\mathcal{G}^{l l l'}_{m -m 0} \propto \delta_{l' 0}$ (see Appendix B of \cite{Komatsu}).
Therefore the projection of the monopole and dipole eliminates all terms from the PDF 
which contain any monopole or dipole modes and importantly does not introduce a new linear
term. After performing this projection the new estimator is
\begin{equation}\label{tSI_suff}
   f_{NL}(\Psi) = \sum_{(l,m)} \mathcal{G}^{l_1 l_2 l_3}_{m_1 m_2 m_3} \frac{\Psi_{l_1 m_1}}{D_{l_1}} 
   \Psi_{l_2 m_2} \Psi_{l_3 m_3},
\end{equation}
where now the sum excludes all modes with $l_i \le 1$.

We must decide whether this estimator contains all the information of the modified PDF. In analogy to the
Poisson case, the $f^2_{NL}$ term is an even function of the $\Psi_{l m}$ modes. Therefore its expectation
value will have a purely Gaussian piece which is fixed by the regularity condition to be precisely the 
correct value for the scale-invariant PDF to satsify Eq. (\ref{nec_suf}) and admit an optimal estimator. 
Again we conclude that $f_{NL}(\Psi)$ is an optimal estimator in the weakly non-Gaussian limit. 

\subsection{Radiative Transfer}

Now we will include the effects of radiative transfer and we will find that the 2-D CMB version
of our optimal estimator is the Weiner Filter estimator used by the WMAP science team
\cite{Komatsu:2003iq}.
The process of radiative transfer alters the amplitudes of the 3-D curvature perturbations
and projects them onto 2-D CMB temperature anisotropies. This process will relate the 
CMB PDF to the Local Model PDF as
\begin{equation}
  P(a|f_{NL}) = \int d^N \Psi \delta^{(M)}(a_{lm} - \int r^2 dr \alpha_l(r) \Psi_{l m}(r)) P(\Psi|f_{NL}),
\end{equation}
where $M < N$ because of the projection. The additional spurious degrees of freedom, which do not affect 
the observable CMB anisotropies, can be integrated out. Here we define 
\begin{equation}
  \alpha_l(r) = \frac{2}{\pi} \int k^2 dk j_l(kr) \Delta_l(k),
\end{equation}
where $\Delta_l(k)$ is the standard radiation transfer function \cite{Ma:1995ey}. We can connect the
present discussion with previous case of the Sachs-Wolfe effect by noting that the formulae of the
previous subsection can be reproduced by choosing 
\begin{equation}
  \alpha_l(r) = \frac{2}{\pi} \int k^2 dk j_l(kr) \frac{1}{3} j_l(k\tau_D).
\end{equation}
By substituting this expression in the following equations the results in the previous subsection can 
be derived. This formula is valid on large angular scales when the wavelength of the relevant primordial curvature 
perturbation is much larger than the size of the sound horizon. 

To calculate this functional integral we will use the exponentiated form of the delta-function
\begin{equation}
   \delta^{(M)}(a_{lm} - \int r^2 dr \Delta_l(r) \Psi_{l m}(r)) 
   = \int d^M B e^{-i\sum_{lm}(-1)^m B_{l -m}(a_{l m} - \int r^2 dr \alpha_l(r) \Psi_{l m}(r)) },
\end{equation}
here $B_{l m}$ is simply a ``dummy variable.'' 
Using this representation we can ``complete the square'' and perform the functional Gaussian integrations.

First we must find the 3-D Local Model PDF, the covariance matrix between two primordial 3-D curvature perturbations
can be calculated as
\begin{eqnarray}
  \mathcal{C} &=& \langle \Phi^{*}_{l_1 m_1}(r_1) \Phi_{l_2 m_2}(r_2) \rangle  
	= \delta_{l_1,l_2} \delta_{m_1,m_2} D_{l_1}(r_1,r_2)	\\
	&=& \delta_{l_1,l_2} \delta_{m_1,m_2} \frac{2}{\pi} \int k^2 dk P(k) j_{l_1}(kr_1) j_{l_1}(kr_2).
\end{eqnarray}
Since we will need the inverse of $\mathcal{C}$ in order to express the PDF of $\Phi$, we can symbolically
define the inverse of $D_{l}(r_1,r_2)$ as
\begin{equation}
  \int r^2 dr D_{l}(r_1,r) D^{-1}_{l}(r,r_2) = \frac{\delta(r_1 - r_2)}{r^2_1}.
\end{equation}
Using these definitions the non-Gaussian PDF can be expanded as
\begin{eqnarray}\label{3D_NG_PDF}
  \log{P(\Psi|f_{NL},\mathcal{C})}&=&\log{P_G(\Psi|f_{NL},\mathcal{C})}+\log{P_{NG}(\Psi|f_{NL},\mathcal{C})} \\  
   &=&-\frac{1}{2} \sum_{l,m} \int r_1^2 dr_1 r_2^2 dr_2 
   (-1)^m \Psi_{l -m}(r_1) D^{-1}_{l}(r_1, r_2) \Psi_{l m}(r_2) \nonumber \\
   &+&f_{NL} \sum_{(l,m)} \int r_1^2 dr_1 r_2^2 dr_2 \mathcal{G}^{l_1 l_2 l_3}_{m_1 m_2 m_3}
   \Psi_{l_1 m_1}(r_1) D^{-1}_{l_1}(r_1 r_2) \Psi_{l_2 m_2}(r_2) \Psi_{l_3 m_3}(r_2) + \mathcal{O}(f^2_{NL}\mu^2) \nonumber,
\end{eqnarray}
where we have ignored the irrelevant constant piece and the linear terms since they will ultimately vanish when 
we project out the monopole mode. 

At this point it will be convenient to introduce the following function
\begin{equation}
  \beta_l(r) = \frac{2}{\pi} \int k^2 dk P(k) j_l(kr) \Delta_l(k).
\end{equation} 
which is defined such that
\begin{equation}\label{ab2c}
   \int r^2 dr \alpha_l(r) \beta_l(r) = C_l.
\end{equation}
Using the definition of $D^{-1}_l(r_1, r_2)$ we can show that
\begin{eqnarray}\label{alpha2beta}
  \alpha_l(r_1) = \int r^2_2 dr_2 D^{-1}_l(r_1, r_2) \beta_l(r_2), \\
  \beta_l(r_1) = \int r^2_2 dr_2 D_l(r_1, r_2) \alpha_l(r_2)
\end{eqnarray}
these properties will be very useful in what follows. Note that the functions $\alpha_l(r)$ and $\beta_l(r)$
are equivalent to $b^{lin}_l(r)$ and $b^{nl}_l(r)$ defined in \cite{Komatsu:2001rj}.

Now we are ready to perform the functional integration and find the non-Gaussian CMB PDF.
We must ``complete the square'' of the following term 
\begin{equation}
   -\frac{1}{2} \sum_{l,m} (-1)^m [\int r_1^2 dr_1 r_2^2 dr_2 (-1)^m \Psi_{l -m} D^{-1}_{l}(r_1, r_2) \Psi_{l m}(r_2) 
   - 2 i  B_{l -m} \int r^2 dr \alpha_l(r) \Psi_{l m}(r)]
\end{equation}
This can be rewritten as
\begin{equation}
   -\frac{1}{2} \sum_{l,m} (-1)^m [\int r_1^2 dr_1 r_2^2 dr_2  (\Psi_{l -m}(r_1) - \zeta_{l -m}(r_1)) 
    D^{-1}_{l}(r_1, r_2) (\Psi_{l m}(r_2) - \zeta_{l m}(r_2) ) 
    + B_{l -m} B_{l m} C_l],
\end{equation}
where
\begin{equation}
\zeta_{l m}(r) = -i\beta_l(r) B_{l m}.
\end{equation}
Once more we must ``complete the square'' for the following term
\begin{equation}
   -\frac{1}{2} \sum_{l,m} (-1)^m [C_l B_{l -m} B_{l m} + 2i B_{l -m} a_{l m}]
\end{equation}
This can be rewritten as
\begin{equation}
   -\frac{1}{2} \sum_{l,m} (-1)^m [C_l (B_{l -m}-\eta_{l -m})(B_{l m} - \eta_{l m}) + \frac{a_{l -m}a_{l m}}{C_l}]
\end{equation}
where
\begin{equation}
\eta_{l m}(r) = \frac{i a_{l m}}{C_l}.
\end{equation}
Thus we are left with the correct Gaussian piece for the CMB PDF.
Performing the two Gaussian functional integrations is equivalent to performing the substitution
\begin{equation}\label{subs}
   \Psi_{l m}(r) \rightarrow \frac{\beta_l(r)}{C_l}a_{l m}.
\end{equation}
into the non-Gaussian cubic portion of the PDF.

This substitution is equivalent to the Weiner Filter solution proposed by \cite{Komatsu:2003iq} in order
to estimate the underlying curvature perturbations from an observed CMB anisotropy map.
The process of projection and radiative transfer is not invertible, but nevertheless we can regularize 
the inversion process by requiring that the reconstructed potential minimizes the variance. This approach 
was adopted in \cite{Komatsu:2003iq}, where it was shown that the optimal inversion procedure was to Weiner 
Filter the CMB map as
\begin{equation}\label{Weiner_Filt}
   \Psi_{l m}(r) = \frac{\beta_l(r)}{C_l} a_{l m}.
\end{equation}
Here we have demonstrated that a straightforward functional integration of the Local Model PDF gives the
same result.

Performing the functional integration, or equivalently substituting the Weiner Filter solution for $\Psi$ 
into Eq. (\ref{3D_NG_PDF}), we find that the non-Gaussian cubic term in the PDF
\begin{eqnarray}
 \log{P_{NG}(a|f_{NL},\mathcal{C})}&=&\sum_{(l,m)} \mathcal{G}_{m_1 m_2 m_3}^{l_1 l_2 l_3} \int r_1^2 dr_1 r_2^2 dr_2 
   \Psi_{l_1 m_1}(r_1) D^{-1}_{l_1}(r_1,r_2)\Psi_{l_2 m_2}(r_2) \Psi_{l_3 m_3}(r_2) \\
    &=& \sum_{(l,m)} \mathcal{G}_{m_1 m_2 m_3}^{l_1 l_2 l_3} \frac{a_{l_1 m_1} a_{l_2 m_2} a_{l_3 m_3}}
    {C_{l_1}C_{l_2}C_{l_3}}
    \int r_1^2 dr_1 r_2^2 dr_2 \beta_{l_1}(r_1) D^{-1}_{l_1}(r_1, r_2)\beta_{l_2}(r_2) \beta_{l_3}(r_2)  \nonumber \\
&=& \sum_{(l,m)'} \mathcal{G}_{m_1 m_2 m_3}^{l_1 l_2 l_3} \frac{a_{l_1 m_1} a_{l_2 m_2} a_{l_3 m_3}}
    {C_{l_1}C_{l_2}C_{l_3}} b_{l_1 l_2 l_3} \nonumber,
\end{eqnarray} 
where we define the reduced CMB Bispectrum as
\begin{equation}\label{lm_cmb_bisp}
   b_{l_1 l_2 l_3} = 2 \int r^2 dr [\alpha_{l_1}(r) \beta_{l_2}(r) \beta_{l_3}(r) + 
      \beta_{l_1}(r) \alpha_{l_2}(r) \beta_{l_3}(r) + \beta_{l_1}(r) \beta_{l_2}(r) \alpha_{l_3}(r)], 
\end{equation}
and the prime indicates that we restrict the sum over $l_i$ such that $l_1 \le l_2 \le l_3$.
It is standard to separate the piece of the CMB three-point correlation function fixed by rotational invariance. 
We do this by introducing the Gaunt Integral, Eq. (\ref{Gaunt}), and the reduced CMB bispectrum, which are 
related to the CMB three-point correlation function as
\begin{equation}
 \langle a_{l_1 m_1} a_{l_2 m_2} a_{l_3 m_3} \rangle = \mathcal{G}_{m_1 m_2 m_3}^{l_1 l_2 l_3} b_{l_1 l_2 l_3}.
\end{equation}
Combining these results we find that the scale-invariant non-Gaussian PDF for the CMB 
anisotropies, ignoring the constant piece, is
\begin{equation}\label{CMB_pdf}
   \log{P(a|f_{NL})} = -\frac{1}{2}\sum_{l,m} \frac{a^{*}_{l m}a_{l m}}{C_l} 
    + f_{NL}\sum_{(l,m)'} \frac{\mathcal{G}_{m_1 m_2 m_3}^{l_1 l_2 l_3}b_{l_1 l_2 l_3}}{C_{l_1}C_{l_2}C_{l_3}} 
    a_{l_1 m_1} a_{l_2 m_2} a_{l_3 m_3} + \mathcal{O}(f^2_{NL}\mu^2).
\end{equation}
Again this has the correct form to admit an optimal estimator, within the weakly non-Gaussian approximation, 
which is
\begin{equation}
\hat{f}_{NL}(a) = \frac{1}{S_{norm}} \sum_{(l,m)'} \frac{\mathcal{G}_{m_1 m_2 m_3}^{l_1 l_2 l_3}b_{l_1 l_2 l_3}}
   {C_{l_1}C_{l_2}C_{l_3}} a_{l_1 m_1} a_{l_2 m_2} a_{l_3 m_3},
\end{equation} 
where the normalization constant is
\begin{equation}
S_{norm} = \sum_{(l,m)'} \frac{(2l_1+1)(2l_2+1)(2l_3+1)}{4\pi}  
    \left(\begin{array}{ccc} l_1 & l_2 & l_3 \\ 0 & 0 & 0 \end{array}\right)^2 
    \frac{b^2_{l_1 l_2 l_3}}{C_{l_1}C_{l_2}C_{l_3}}.
\end{equation}

A procedure based on this estimator has been implemented by the WMAP science team in their analysis 
of primordial non-Gaussianity. Their interpretation of the estimator is in terms of Weiner Filtered
estimates of the primordial curvature. First the Weiner Filter $\beta_l(r)/C_l$ is used to estimate the 
primordial curvature fluctuation $\Psi_{lm}(r)$ from the CMB anisotropy $a_{lm}$, as in Eq. (\ref{Weiner_Filt}). 
Then the two estimates of the primordial curvature perturbations are used to estimate the quadratic piece 
of the primordial curvature field according to the definition of the Local Model and $\alpha_l(r)$ is 
used to calculate the CMB temperature anisotropy due to this quadratic piece. This quadratic CMB 
template is correlated with the observed CMB anisotropies to give an estimate, once properly normalized, 
of the non-Gaussianity of the observed CMB anisotropies. We have shown that this intuitive procedure results 
in an estimator that is optimal.  

\section{General non-Gaussian Models}

In the previous section we applied the basic ideas of estimation theory to show that the Local Model
non-Gaussian CMB PDF admitted an optimal estimator in the weakly non-Gaussian limit. As mentioned
in the Introduction the Local Model is a physically motivated idealization, however most primordial
bispectra calculated within models of the early universe contain additional terms. Starting with
the definition of the Local Model, Eq. (\ref{local}), we derived the resulting CMB PDF. Since it
is not possible to find the general non-Gaussian PDF which corresponds to a general non-Gaussian
bispectrum, we are unable to retrace the above steps in order to extend our analysis. However we 
will argue that the non-Gaussian CMB PDF in Eq. (\ref{CMB_pdf}) is simply the Edgeworth expansion, 
which holds for arbitrary non-Gaussian CMB PDF, and therefore our conclusion that the bispectrum 
estimator is optimal holds for arbitrary model of the primordial non-Gaussianity. Due to the arguments 
in the Introduction we assume that a model for primordial non-Gaussianity is best characterized by 
its bispectrum and that we should try to constrain the amplitude of the bispectrum. We choose to define the 
amplitude of a general bispectrum as the coefficient of the bispectrum evaluated in the equilateral 
configuration (see \cite{Babich:2004gb} for a full discussion of this point). 

The Edgeworth expansion is a way to express the non-Gaussianity of a PDF in the form of a series 
expansion (see \cite{bernardeau} and references there within). This allows one to explicitly 
write down a non-Gaussian PDF if its lowest order moments or correlation functions are known. 
The Edgeworth expansion of a 1-D PDF is a simple expansion in Hermite polynomials;
a multivariate generalization has been found \cite{Taylor:2000hq, Bernardeau:1994aq}. 
Adopting the notation relevant for CMB anisotropies the Edgeworth expansion is
\begin{equation}
  P(a) = [1 - \sum_{(l,m)'} \langle a_{l_1 m_1} a_{l_2 m_2} a_{l_3 m_3} \rangle 
   \frac{\partial}{\partial a_{l_1 m_1}} \frac{\partial}{\partial a_{l_2 m_2}} \frac{\partial}{\partial a_{l_3 m_3}}]
   \prod_{l m} \frac{ e^{ -\frac{ a^{*}_{l m} a_{l m} } {2C_l}} } {\sqrt{2\pi C_l}}, 
\end{equation}
where $\langle a_{l_1 m_1} a_{l_2 m_2} a_{l_3 m_3} \rangle$ can still be decomposed into the
Gaunt Integral and the reduced bispectrum. Here the general reduced bispectrum can be 
calculated as
\begin{equation}
   b_{l_1 l_2 l_3} = (\frac{2}{\pi})^3 \int k_1^2 dk_1 k_2^2 dk_2 k_3^2 dk_3 \int x^2 dx 
   j_{l_1}(k_1 x) j_{l_2}(k_2 x)j_{l_3}(k_3 x)
   B(k_1, k_2, k_3) \Delta_{l_1}(k_1) \Delta_{l_2}(k_2) \Delta_{l_3}(k_3),
\end{equation}
where $B(k_1, k_2, k_3)$ is the general primordial bispectrum \cite{Babich:2004gb}. 
When the primordial bispectrum is calculated according to the Local Model the CMB reduced 
bispectrum is given by Eq. (\ref{lm_cmb_bisp}).

Performing the functional differentiation in the Edgeworth expansion we find
\begin{eqnarray}
  P(a)&=&\prod_{l m} \frac{ e^{ -\frac{ a^{*}_{l m} a_{l m} } {2C_l}} } {\sqrt{2\pi C_l}} [1 
   + \sum_{(l,m)'}  b_{l_1 l_2 l_3} \mathcal{G}^{l_1 l_2 l_3}_{m_1 m_2 m_3}   
   [\frac{a_{l_1 m_1} a_{l_2 m_2} a_{l_3 m_3}}{C_{l_1} C_{l_2} C_{l_3}} \nonumber \\
   &-& (-1)^{m_3}( \frac{a_{l_1 m_1}}{C_{l_1}C_{l_3}}\delta_{l_2,l_3} \delta_{m_2,-m_3} +
   \frac{a_{l_2 m_2}}{C_{l_2}C_{l_3}}\delta_{l_1,l_3} \delta_{m_1,-m_3} +	
   \frac{a_{l_3 m_3}}{C_{l_2}C_{l_3}}\delta_{l_1,l_2} \delta_{m_1,-m_2} )]].
\end{eqnarray}
Again the properties of the Gaunt Integral force the linear terms to be proportional to
the monopole anisotropy mode, $a_{0 0}$. After projecting out the monopole anisotropy 
the form of the non-Gaussian cubic term in the Edgeworth expansion is identitical to 
the form of non-Gaussian cubic term in the Local Model CMB PDF, Eq. (\ref{CMB_pdf}).

In the previous section we argued that estimators based on the bispectrum are optimal if the
PDF can be expressed as in Eq. (\ref{CMB_pdf}). This form is not unique to the Local 
Model, but is simply part of the Edgeworth expansion which is relevant regardless the form of
primordial bispectrum. The only necessary condition is that the level of non-Gaussianity is
small. Thus we conclude that estimators based on the bispectrum are optimal for any form of 
primordial non-Gaussianity characterized by its bispectrum provided the amplitude of the 
non-Gaussianity is sufficiently small. 

\section{Discussion}

We analyzed the standard model for primordial non-Gaussianity with the tools of estimation 
theory and found that the estimator constructed out of the CMB three-point correlation function 
is an optimal estimator for $f_{NL}$, the amplitude of the primordial non-Gaussianity. 
Our conclusion is only true within the weakly non-Gaussian approximation, which implies that 
we ignore non-Gaussian contributions to four-point correlation function compared to the 
much larger Gaussian contributions. In our calculations this is equivalent to ignoring 
terms proportional to $f^2_{NL} \langle \Phi({\bm x})^2\rangle$, which is already 
constrained by the WMAP data to be extremely small. Therefore we can consider the standard 
WMAP estimator, which is based on the CMB three-point correlation function, to be optimal 
in practice. This property only depends on the weak level of non-Gaussianity, not on any 
assumed form for the primordial bispectrum. Therefore we argued that the amplitude of a 
general primordial bispectrum can be optimally estimated. 

Our calculations demonstrated that the WMAP estimator was optimal to constrain $f_{NL}$ and
therefore contained all the information in the observed data on this parameter. Future work 
can now focus on practical implementations of this optimal estimator. Some of the practical 
concerns that affect the implementation of an estimator for $f_{NL}$ include biasing due to 
non-Gaussianity from secondary anisotropies, non-uniform instrument noise and the need to 
jointly estimate $f_{NL}$ and the basic cosmological parameters from the same data which will 
degrade the performance of the estimator. While we demonstrated that the WMAP estimator is 
optimal for the estimation of $f_{NL}$, there still is a need to find a quick method to optimally 
estimate individual three-point correlation function modes from CMB maps with Galactic foreground 
contamination and inhomogeneous instrument noise \cite{Heavens:1998jb,Santos:2002df}. We implicitly 
assumed that the three-point correlation function modes could simply
be calculated from observed CMB maps. For maps with inhomogeneous noise, this procedure results
in sub-optimal error bars on the individual bispectra modes. An optimal procedure, which is quite
slow for large data sets, has been developed \cite{Heavens:1998jb}; a quicker procedure must be
found if it can be realistically applied to the WMAP and Planck data sets.

Also, the non-Gaussianity of secondary anisotropies is not well known on the arcminute scales 
relevant for current and future CMB experiments. Fortunately on intermediate scales there is 
little contamination of the primordial non-Gaussianity estimator by secondary anisotropies; 
however we simply do not know if this is true on the extremely small scales relevant for 
upcoming CMB experiments. Also we do not know how errors in the basic cosmological parameter, 
from which we calculate the radiation transfer functions and the Weiner Filters, will affect
this analysis. This topic should also be investigated.

While much progress has been made in both the theoretical 
understanding and practical analysis of the signatures of primordial non-Gaussianity in CMB maps, 
there still is much more work needed to be done before the field will become fully developed. 
However the tremendous potential for insight into the production mechanism of the primordial 
curvature perturbations makes this work worthwhile.

\begin{acknowledgments}
   We would like to thank Oliver Zahn, Chris Hirata and Antony Lewis for useful conversations
and especially Matias Zaldarriaga for a careful reading of the manuscript and many helpful 
conversations.
\end{acknowledgments}

\appendix

\section{Signal-to-Noise Ratio of Higher-Order Estimators}
In this Appendix we demonstrate the features of the $S/N$ of an estimator based on 
an $n^{th}$-order correlation function needed to argue that the estimator based
on the three-point correlation function would have the largest $S/N$. 
We will do our calculation within a toy model that should retain all of the important features
of the problem. In what follows we ignore radiative transfer and the curvature of the sky, so 
we will simply observe the underlying modes within the flat-sky approximation \cite{babich2}.

Extending the basic idea of the ``Local Model'' we will assume that the observed non-Gaussian 
field $\Psi$ can be expanded in the Gaussian field $\Phi$ as
\begin{equation}\label{elm}
  \Psi = \Phi + f_2 [\Phi^2 - \langle \Phi^2 \rangle] + f_3 [\Phi^3 - \langle \Phi^3 \rangle] +
   \cdots + f_n [\Phi^n - \langle \Phi^n \rangle] + \cdots.
\end{equation}
The various n-point correlations function of $\Psi$ can be calculated by substituting our definition of
$\Psi$, Eq. (\ref{elm}), and using Wick's Theorem. It is important to note that we are interested
in connected correlation functions, which contain a single delta-function. In what follows we 
ignore all $\mathcal{O}(1)$ factors.

Working within our model, the power spectrum is evaluated as \cite{babich2}
\begin{equation}
  C(l) \sim \frac{A}{l^2}, 
\end{equation}
and the connected $n^{th}$-order correlation function as
\begin{equation}
   T(\bm{l}_1,\bm{l}_2, \cdots, \bm{l}_n) \sim (f_{n-1} + f^2_{n-2} + \cdots + f^{n-s}_{s} + \cdots f^{n-2}_2) 
   A^{n-1} \frac{l^2_1 + l^2_2 + \cdots + l^2_n}{l_1^2 l_2^2 \cdots l_n^2}, 
\end{equation}
where $A$ is the amplitude of the primordial curvature power spectrum, $P(k) = A/k^3$. Note that there is
no contribution to $n^{th}$-order correlation function from $f_{n}$.

We also define the total $(S/N)^2$ of the estimator as
\begin{equation}\label{sn_def}
(\frac{S}{N})^2 \sim \int d^2\bm{l}_1 d^2\bm{l}_2 \cdots d^2\bm{l}_n 
  \frac{[\delta^{(2)}(\bm{l}_1 + \bm{l}_2 + \cdots + \bm{l}_n) T(\bm{l}_1,\bm{l}_2, \cdots, \bm{l}_n)]^2}
  {C(l_1)C(l_2) \cdots C(l_n)}
\end{equation}
Substituting the results for $C_l$ and $T(\bm{l}_1,\bm{l}_2,\cdots,\bm{l}_n)$ into Eq. (\ref{sn_def}) we find
\begin{equation}
(\frac{S}{N})^2 \sim f_{sky} f_{tot}^2 
 A^{n-2} \int d^2\bm{l}_1 d^2\bm{l}_2 \cdots d^2\bm{l}_n 
\delta^{(2)}(\bm{l}_1 + \bm{l}_2 + \cdots + \bm{l}_n) \frac{[l_1^2 + l_2^2 + \cdots + l_n^2]^2}
 {l_1^2 l_2^2 \cdots l_n^2},
\end{equation}
where $f_{sky}$ is the fraction of the observed sky and $f_{tot} 
= f_{n-1} + f^2_{n-2} + \cdots + f^{n-s}_{s} + \cdots f^{n-2}_2 $ is the total coupling cofficient 
that results from considering all contribution to a given correlation function.

The upper bound of integration for all $l_i$ is $l_{max}$ so we can rescale the integral to be
``dimensionless,'' then we find
\begin{equation}
 (\frac{S}{N})^2 \sim f_{sky} f_{tot}^2 A^{n-2} l^2_{max} = f_{tot}^2 A^{n-2} N_{pix},
\end{equation}
where we defined $N_{pix}$ to be the total number of observed pixels. In addition to numerical 
factors, the integral may also contribute factors of the ``Coulomb Logarithm,'' i.e. $\ln(l_{max}/l_{min})$, 
as found in the bispectrum calculation \cite{babich2}. These additional correction are small, so we
can argue that to $\mathcal{O}(1)$, the $S/N$ is simply the product of the characteristic amplitude
and the square-root of the number of observed pixels. We have established the necessary results 
used in the Introduction where we argued that the $S/N$ of the estimator constructed out of 
the three-point correlation functions will be dominant.

\end{document}